\documentclass[aps,prl,superscriptaddress,twocolumn]{revtex4}
\usepackage{graphicx}
\usepackage{epstopdf}
\usepackage{bm}
\begin{document}

\title{Two-stage ordering of spins in a dipolar spin ice on the kagome lattice}
\author{Gia-Wei Chern}
\affiliation{Department of Physics, University of Wisconsin, Madison, Wisconsin 53706, USA}
\author{Paula Mellado}
\affiliation{School of Engineering and Applied Sciences, Harvard University, Cambridge, Massachusetts 02138, USA}
\author{O. Tchernyshyov}
\affiliation{Department of Physics and Astronomy, Johns Hopkins University, Baltimore, Maryland 21218, USA}

\begin{abstract}
Spin ice, a peculiar thermal  state of a frustrated ferromagnet on
the pyrochlore lattice, has a finite entropy density and excitations
carrying magnetic charge.  By combining analytical arguments and
Monte Carlo simulations, we show that spin ice on the two-dimensional kagome
lattice orders in two stages. The intermediate phase has ordered magnetic charges
and is separated from the paramagnetic phase by an Ising transition. The transition to
the low-temperature phase is of the three-state Potts or Kosterlitz-Thouless type,
depending on the presence of defects in charge order. 
\end{abstract}

\maketitle

Frustrated magnets \cite{IFM} attract
attention of both  theorists and experimentalists as models of
strongly interacting systems with unusual ground states and
elementary excitations.  One of the recent surprises was a
realization that elementary excitations in spin ice
\cite{gingras.review.2009} are quasiparticles
carrying magnetic charge \cite{Nature.451.42}.  Subsequent work
\cite{NatPhys.5.258} uncovered signatures of the magnetic monopoles
in magnetization dynamics of spin ice.  Spin ice is a frustrated
ferromagnet discovered in the pyrochlore Ho$_2$Ti$_2$O$_7$,  where
magnetic Ho$^{3+}$ ions form a network of corner-sharing tetrahedra
\cite{PhysRevLett.79.2554}.  The magnetic moments $\bm \mu_i = 
\sigma_i \mu \hat \mathbf e_i = \pm
\mu \hat \mathbf e_i$ are forced to point along a $\langle 111
\rangle$ axis $\hat \mathbf e_i$ by a strong crystal field.  The
easy-axis anisotropy makes the spins Ising-like, so that a
microstate of this magnet can be described by Ising variables
$\sigma_i = \pm 1$.  The Hamiltonian of spin ice includes exchange interactions
of strength $J$ for pairs of nearest neighbors $\langle ij \rangle$
and dipolar interactions between all spins \cite{gingras.review.2009}:
\begin{eqnarray}
H &=& -J\sum_{\langle ij \rangle} \sigma_i \sigma_j
    (\hat \mathbf e_i \cdot \hat \mathbf e_j)
\label{eq:H}\\
&&+ \frac{Dr_\mathrm{nn}^3}{2} \sum_{i \neq j} \sigma_i \sigma_j
    \frac{(\hat \mathbf e_i \cdot \hat \mathbf e_j) - 3(\hat \mathbf e_i \cdot \hat \mathbf r_{ij})(\hat \mathbf e_j \cdot \hat \mathbf r_{ij})}{|\mathbf r_i -\mathbf r_j|^3},
    \nonumber
\end{eqnarray}
where $D = (\mu_0/4\pi)\mu^2/r_\mathrm{nn}^3$ is a characteristic
strength of dipolar coupling, $\mathbf r_i$ are spin locations,
$\hat \mathbf r_{ij} = (\mathbf r_i - \mathbf r_j)/|\mathbf r_i -
\mathbf r_j|$, and $r_\mathrm{nn}$ is the distance between nearest
neighbors.  In the absence of dipolar interactions, $D=0$, and for
ferromagnetic exchange, $J>0$, the system is strongly frustrated
because it is impossible to minimize the energy of every bond
$\langle ij \rangle$.  In a ground state, two spins point into every
tetrahedron and two point out, which is reminiscent of proton
positions in water ice, where every oxygen has
two protons nearby and two farther away.   This ice rule is
satisfied by a macroscopically large number of microstates, so that
both protons in water ice and magnetic moments in spin ice can
remain disordered even at low temperatures \cite{nature.399.333}.

Large magnetic moments ($\mu = 10 \mu_B$ in Ho$_2$Ti$_2$O$_7$) make
magnetic dipolar interactions between nearest neighbors comparable
to exchange \cite{PhysRevLett.84.3430}.  Together
with the long-distance nature of dipolar interactions, the
substantial value of $D$ casts doubt on the usefulness of the
short-range ($D=0$) model of spin ice.  Yet numerical simulations
show that, even after the inclusion of dipolar interactions, energy
differences between states obeying the ice rule remain numerically
small---so small that magnetic order induced by the dipolar
interactions is expected to occur only at a rather low temperature,
$T \approx 0.13 D$ \cite{PhysRevLett.87.067203, CanJPhys.79.1339, PhysRevLett.95.217201}. 
The persistent near-degeneracy of ice ground states in the presence 
of dipolar interactions was clarified by Castelnovo
\textit{et al.} \cite{Nature.451.42}, who introduced a
``dumbbell" version of spin ice, in which magnetic dipoles
are stretched into bar magnets of length $a$ such that
their poles meet at the centers of tetrahedra.  The energy of the resulting model can
be represented as a Coulomb interaction of magnetic charges 
of the dumbbells, $q_i = \pm \mu/a$ \cite{Nature.451.42}:
\begin{equation}
E(\{Q_i\}) = \sum_{\alpha} \frac{Q_\alpha^2}{2C}
    + \frac{\mu_0}{8\pi}\sum_{\alpha \neq \beta}
    \frac{Q_\alpha Q_\beta}{|\mathbf r_\alpha - \mathbf r_\beta|}.
\label{eq:E}
\end{equation}
In this expression, $Q_\alpha = \sum_{i \in \alpha} q_i$ is the sum
of  magnetic charges at the center of tetrahedron $\alpha$.  In a
spin-ice state of the dumbbell model, every tetrahedron has
two north and two south poles with a total magnetic charge
$Q_\alpha=0$, minimizing the first term in Eq.~(\ref{eq:E}).  As a
result, no magnetic field will be generated and the magnetic dipolar
energy is strictly zero.  A partial cancellation occurs in the original model 
(\ref{eq:H}), making the Coulomb energy (\ref{eq:E}) a very good approximation.  
The charge of tetrahedron $\alpha$, expressed in units of $\mu/a$, is
\begin{equation}
Q_\alpha = \pm \sum_{i \in \alpha} \sigma_i,
\label{eq:Q}
\end{equation}
with the plus sign for one sublattice of tetrahedra and minus for
the other.   Residual interactions, responsible for the formation of
magnetic order, are weak and fall off quickly with the distance
\cite{Nature.451.42}.  The resulting energy differences between
states obeying the ice rule are only a small fraction of the dipolar
energy scale $D$ \cite{CanJPhys.79.1339, PhysRevLett.95.217201, JPCM.16.R1277}.  As the
magnet is cooled down from a high-temperature paramagnetic state
with completely uncorrelated spins, it first gradually enters the
spin-ice regime at the crossover temperature $T \approx
2J_\mathrm{eff}$, where $J_{\rm eff} = J/3 + 5D/3$ is  the effective
interaction for nearest-neighbor Ising spins $\sigma_i$
\cite{gingras.review.2009}, and then undergoes a
phase transition into a magnetically ordered state at $T \approx
0.13 D$.

\begin{figure}
\includegraphics[width=0.97\columnwidth]{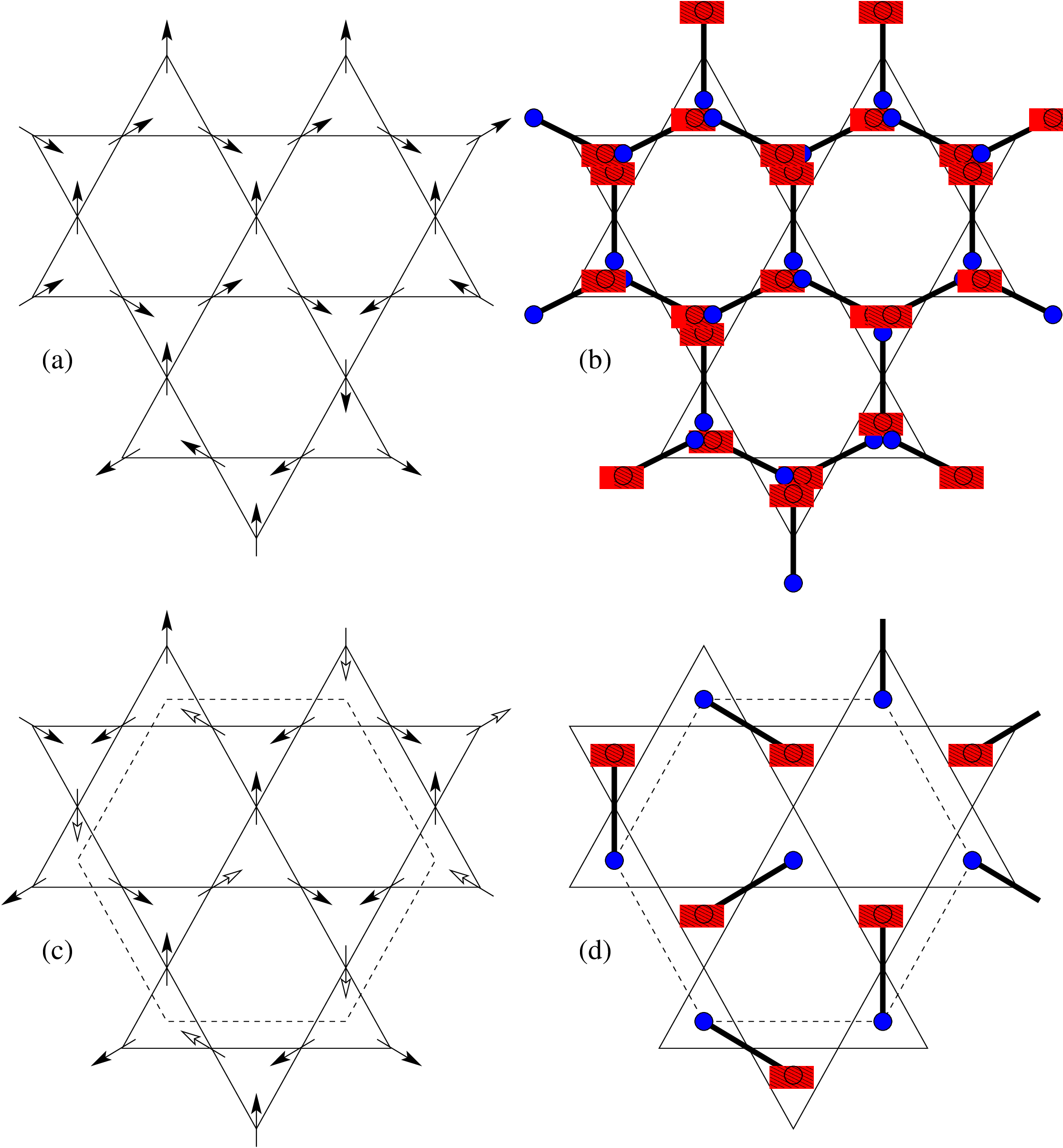}
\caption{Magnetic configurations of the dipolar kagome ice and their alternative representations.  
(a) A spin-ice microstate lacking spin order but possessing charge order.  
The latter is manifested in the dumbbell representation (b). (c) One of the ground states exhibiting 
the $\sqrt{3} \times \sqrt{3}$ magnetic order and its depiction in terms of dimers (d).  
The dashed line marks the magnetic unit cell.}
\label{fig:dipoles-bars}
\end{figure}

In this Letter, we discuss a related problem of dipolar spin ice on
kagome, a two-dimensional lattice of corner-sharing triangles.  Each
spin is constrained to point along the line connecting the two
triangles, Fig.~\ref{fig:dipoles-bars}(a).  Possible ways of
realizing such a system are discussed below.   A short-range version
of this model ($D=0$, $J>0$) was studied by Wills \textit et al.
\cite{PhysRevB.66.144407}.

At a first glance, the
dipolar spin ice on kagome is very similar to its counterpart on the
pyrochlore lattice and one might expect a similar sequence of
transformations, namely a crossover to a correlated but disordered spin-ice state
followed by a transition into a magnetically ordered phase. However, a closer
look reveals the existence of an intermediate thermodynamic phase
with ordered magnetic charges and disordered spins. To see this, note
that the allowed values of magnetic charge (\ref{eq:Q}) are
even on a tetrahedron and odd on a triangle. Consequently, minimization of the
first term in Eq.~(\ref{eq:E}) yields $Q_\alpha = 0$ on the
pyrochlore lattice and $ \pm 1$ on kagome.  A microstate obeying the
ice rule $Q_\alpha = \pm 1$, shown in
Figs.~\ref{fig:dipoles-bars}(a) and (b), contains uncompensated
charges that generate a magnetic field.  To a first approximation,
the system energy is given by the Coulomb term in (\ref{eq:E}).
Nonzero values of magnetic charges result in substantial energy
differences between states obeying the ice rule.  The
Coulomb energy is minimized when adjacent triangles carry charges of opposite signs.  
The charge-ordered states of the dipolar kagome ice are closely related
to the ice states of the pyrochlore spin ice in a $\langle 111 \rangle$ 
magnetic field \cite{JPSJ.71.2365}.  The number
of such states grows exponentially with the number of spins $N$.
They are exactly degenerate in the dumbbell model.  In the dipolar
ice model (\ref{eq:H}), the degeneracy is lifted by small corrections to the
Coulomb energy (\ref{eq:E}).

This hierarchy of energy scales suggests the following sequence of
thermal transformations in the dipolar spin ice on kagome.  As the
magnet cools down from the high-temperature paramagnetic state with
entropy per spin $s = \ln{2} = 0.693$, it gradually enters a
spin-ice state with $s \approx (1/3)\ln{(9/2)} = 0.501$
\cite{PhysRevB.66.144407}.  At a temperature of the order of $D$, it
will enter a distinct phase with ordered magnetic charges, where entropy density 
is reduced but remains nonzero, $s = 0.108$ per spin; the state is similar to that of the 
pyrochlore spin ice in a magnetic field along $\langle 111 \rangle$ \cite{JPSJ.71.2365}.  
At an even lower temperature, the system will enter a spin-ordered state with zero entropy density.
In contrast, spin ice with nearest-neighbor interactions only exhibits neither charge, 
nor spin order \cite{PhysRevB.66.144407}.

\begin{figure}
\includegraphics[width=0.99\columnwidth]{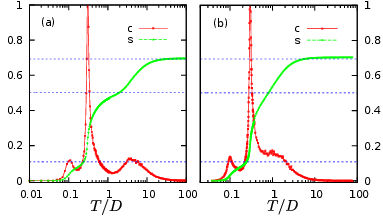}
\caption{Temperature dependence of the specific heat $c(T)$ and
entropy per spin $s(T)$ of the dipolar spin ice (\ref{eq:H}) with
(a) ferromagnetic exchange $J = 0.5 D$ and (b) antiferromagnetic
exchange $J=-2.67 D$. The linear size of the system is $L=12$. The
dashed lines show levels of entropy $s = 0.693$ (Ising paramagnet), $0.501$ (spin ice), 
and $0.108$ (charge-ordered spin ice) per spin.}
\label{fig:results}
\end{figure}

The above scenario of two-stage spin ordering is confirmed by our Monte Carlo 
simulations on the kagome dipolar ice model (\ref{eq:H}). A similar
conclusion was reached independently in Ref.~\cite{moller}.
The specific heat $c(T)$ and entropy per spin $s(T)$ are shown in
Fig.~\ref{fig:results} for ferromagnetic exchange $J =0.5 D$, and
antiferromagnetic $J = -2.67 D$. In both cases, a broad peak in
$c(T)$ signals a crossover from the paramagnetic regime to the
spin-ice state.  The latter is seen as a washed-out plateau in
$s(T)$ near the characteristic spin-ice value $s \approx 0.501$. The
crossover temperature $T \approx 2J_{\rm eff}$, where $J_{\rm eff} =
J/2 + 7D/4$ on kagome. Two sharp peaks in $c(T)$ at lower
temperatures mark the charge and spin-ordering phase transitions. In
the antiferromagnetic case, $J = -2.67 D$, the effective
nearest-neighbor Ising coupling becomes small, $J_{\rm eff} = 0.415
D$, so that the system is near the spin ice--antiferromagnet
boundary, given approximately by the condition $J_{\rm eff} = 0$
\cite{gingras.review.2009}.  This makes the spin-ice plateau
indistinct, but the two phase transitions are clearly present.

\begin{figure}
\includegraphics[width=0.99\columnwidth]{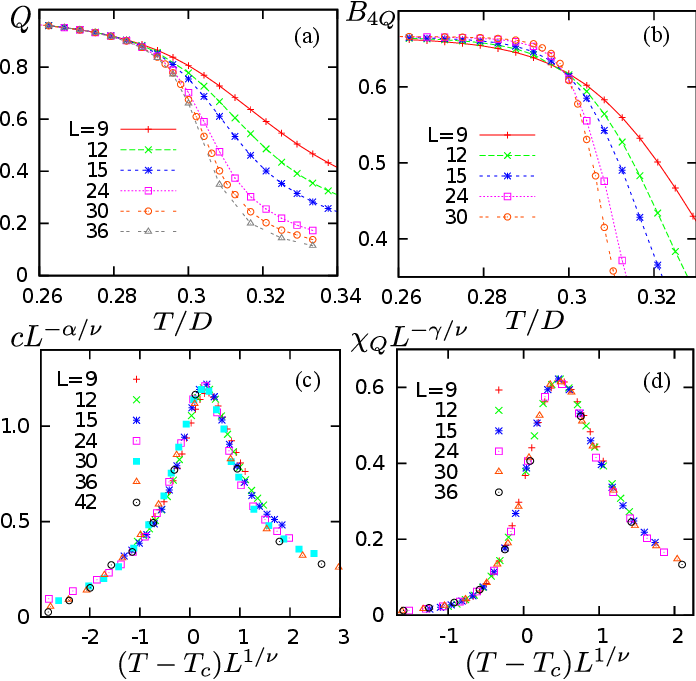}
\caption{Monte Carlo simulation of the charge-ordering transition in the spin-ice model (\ref{eq:H}).  
(a) and (b) show the temperature dependence of the staggered charge order parameter $Q$ and
Binder's fourth-order cumulant $B_{4Q} \equiv 1-\langle
Q^4\rangle/3\langle Q^2\rangle^2$. (c) and (d) show the scaling of specific heat $c$ and charge-order susceptibility $\chi_Q= 
(\langle Q^2 \rangle - \langle Q \rangle^2)/NT$
using critical exponents $\alpha = 0$, $\gamma = 7/4$ and $\nu = 1$ from the 2D Ising universality class.}
\label{fig:charge-order}
\end{figure}

We first discuss the details of the charge ordering transition 
and focus on the case of ferromagnetic exchange $J = 0.5D$. Monte Carlo simulations
were performed with periodic boundary conditions and linear sizes up to $L=36$, or $N = 3L^2 = 3888$ sites. Long-range dipolar interactions are summed over periodic copies up to a distance
of $500 L$. The temperature dependence of the staggered charge order parameter $Q$ is shown 
in Fig.~\ref{fig:charge-order}(a) for various system sizes. The existence of a continuous 
charge-ordering transition is evidenced by Fig.~\ref{fig:charge-order}(b), where
Binder's fourth-order cumulants of different $L$ cross at $T_c \approx 0.267\,D$.

The $Z_2$ symmetry of the order parameter $Q$ suggests that the charge-ordering
transition is in the universality class of the Ising model. To verify this conjecture, 
we performed a finite-size scaling analysis and found excellent data collapse
with the critical exponents of the 2D Ising universality class. 
Figs.~\ref{fig:charge-order}(c) and (d) show the scaling behavior of 
the specific heat $c$ and charge susceptibility $\chi_Q$. In obtaining the scaled curves,
we have subtracted a size-dependent background contribution from
the specific-heat.

The spin order emerging on top of magnetic charge order is expected
to be that of $\sqrt{3} \times \sqrt{3}$ type shown in
Fig.~\ref{fig:dipoles-bars}(c), the same as in the short-range model
with antiferromagnetic second-neighbor exchange
\cite{PhysRevB.66.144407}.  This can be understood as follows.  In a
charge-ordered state every triangle has two majority spins pointing
into (or out of) the triangle and a minority spin pointing the other
way.  Such states can be represented by dimer coverings of a
honeycomb lattice, Fig.~\ref{fig:dipoles-bars}(d); the dimers
indicate locations of minority spins.  The energy of such a state is
determined by the interactions between minority spins alone.  To see
that, picture a minority spin $-\bm \mu$ as a superposition of a
majority spin $+\bm \mu$ and a minority spin of double strength $-2
\bm \mu$;  in this representation, majority spins form an inert
background.  We thus arrive at a model of dimers with point dipoles
of strength $2\mu$ directed along the dimers and towards triangles
with positive charge.  The interaction energy of two dimers depends
on their mutual position.  It is minimized by increasing the number
of second neighbors (distance between centers $\sqrt{3}
r_\mathrm{nn}$) and reducing the number of third neighbors ($2
r_\mathrm{nn}$).  The dimer configuration shown in
Fig.~\ref{fig:dipoles-bars}(d) optimizes both.  It is one of three
states related to each other by lattice translations.    The
corresponding spin order is shown in Fig.~\ref{fig:dipoles-bars}(c).
There are a total of 6 magnetic ground states related to each other
by lattice translations and time reversal.

\begin{figure}[t]
\includegraphics[width=0.95\columnwidth]{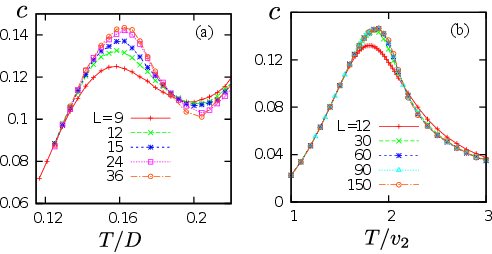}
\caption{Specific heat $c$ as a function of temperature in Monte Carlo simulations
of (a) dipolar spin ice model Eq.~(\ref{eq:H}) on kagome, and (b) dimer model with attractive $v_2$
on honeycomb lattice. The peak of the specific heat curve corresponds to $\sqrt{3}\times\sqrt{3}$
magnetic and dimer ordering.}
\label{fig:c-t}
\end{figure}

The symmetry-breaking pattern of the magnetic order described above
suggests that the magnetic transition is in the universality class
of the 2D three-state Potts model. However, Monte Carlo simulations of the dipolar ice model on systems
up to $L=36$ fail to turn up any evidence of the Potts critical behavior.  The lack of a singularity in the specific heat
[Fig.~\ref{fig:c-t}(a)] is consistent with a Kosterlitz--Thouless (KT) transition.
This unexpected result can be understood by exploiting the previously mentioned mapping 
between ice states with perfect charge order and dimer coverings of the honeycomb lattice.
The minimal dimer model consistent with the required magnetic order includes attraction 
between second-neighbor dimers $v_2$. We performed large-scale Monte Carlo simulations for
the honeycomb dimer model using the worm algorithm \cite{sandvik}. As shown
in Fig.~\ref{fig:c-t}(b), the transition into the 
$\sqrt{3}\times\sqrt{3}$ dimer order is indeed characterized by a KT transition similar 
to the case of dimer model on square lattice \cite{alet}.

\begin{figure}[t]
\includegraphics[width=0.97\columnwidth]{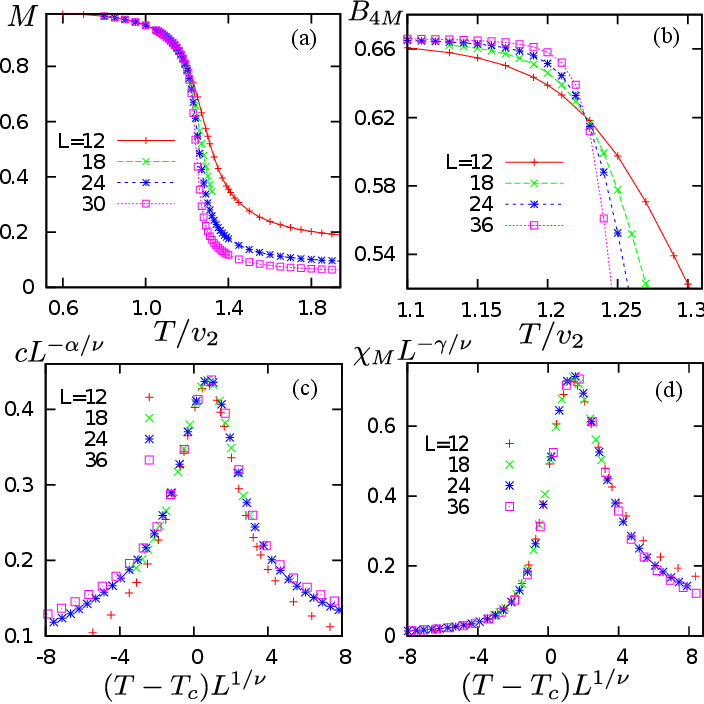}
\caption{Monte Carlo simulations of dimer ordering on a honeycomb lattice with
a finite fugacity for charge defects. 
(a) and (b) show the temperature dependence of the magnetic
order parameter $M$ and Binder's fourth-order cumulant $B_{4M}
= 1-\langle |M|^4\rangle/3\langle |M|^2\rangle^2$.  (c) and (d) show the scaled specific heat $c$ and magnetic susceptibility
$\chi_M = (\langle |M|^2 \rangle - |\langle M
\rangle|^2)/NT$. The critical exponents of 2D three-state Potts
model $\alpha = 1/3$, $\gamma = 13/9$, and $\nu = 5/6$ are used.}
\label{fig:magnetic-order}
\end{figure}

The presence of thermally excited defects in charge order spoils the mapping to dimer coverings. The counterpart of a charge defect in the honeycomb dimer model is a site with 
two dimers attached to it.  We performed simulations with a finite fugacity 
$z = \exp(-\varepsilon/T)$ for such defects. The results for $\varepsilon = 2v_2$ are shown 
in Fig.~\ref{fig:magnetic-order}. The corresponding magnetic order parameter $M$ is shown 
as a function of temperature in Fig. \ref{fig:magnetic-order}(a). Binder's cumulants cross at $T_m \approx 1.226\,v_2$
[Fig. \ref{fig:magnetic-order}(b)], indicating a continuous phase transition. 
Finite-size scaling with the critical exponents of the three-state
Potts model gives excellent data collapse, Figs.~\ref{fig:magnetic-order}(c) and (d).

These results illustrate the importance of charge defects for magnetic ordering. When 
the average separation between charge defects exceeds the system size, spin correlations 
decay algebraically with the distance, as expected from the dimer mapping. Consequently,
one observes a KT-like magnetic transition for small systems of the dipolar ice model. 
The critical behavior characteristic of the three-state Potts universality only reveals 
itself for sufficiently large spin systems.

In summary, we presented a plausible scenario of a two-stage magnetic ordering in the dipolar spin ice on kagome.  In contrast to spin ice on the pyrochlore lattice, this model has an intermediate phase distinguished by an Ising order of magnetic charges. The transition to 
the low-temperature phase is of the Kosterlitz-Thouless type if defects
in charge order are absent and of the three-state Potts type if they are present.
Kagome spin ice already exists as an artificial magnetic lattice \cite{tanaka:052411,qi:094418}.  Although the energy scale of dipolar interactions in artificial ice greatly exceeds the room temperature, it may be possible to introduce thermal motion of spins by agitating them in a manner of granular systems \cite{ke:037205}.

We thank John Cumings, Roderich Moessner, and Peter Schiffer for valuable discussions.  This work was supported by the NSF Award DMR-0348679 and by the DOE Award DE-FG02-08ER46544.

\end{document}